\documentclass{appolb}
\usepackage{graphicx}
\usepackage[T1]{fontenc} 
\usepackage{graphicx}
\usepackage{dcolumn}
\usepackage{bm}
\usepackage{slashed}
\usepackage{amsmath,graphicx}
\usepackage[colorlinks=true,linktocpage=true,linkcolor=blue,citecolor=blue]{hyperref}
\usepackage{float}
\usepackage{nicefrac}
\usepackage[normalem]{ulem}
\usepackage{subfigure} 
\usepackage{bbold} 
\usepackage[makeroom]{cancel}

\def\be{\begin{equation}}
\def\ee{\end{equation}}
\newcommand{\bel}[1]{\begin{eqnarray}\label{#1}}
\newcommand{\eel}{\end{eqnarray}}
\def\barr{\begin{array}}
	\def\earr{\end{array}}
\def\beq{\begin{eqnarray}}
\def\eeq{\end{eqnarray}}
\def\bfig{\begin{figure}}
	\def\efig{\end{figure}}

\newcommand{\f}[2]{\frac{#1}{#2}}
\newcommand{\onehalf}{{\nicefrac{1}{2}}}

\newcommand{\p}{\partial}

\newcommand{\tr}{{\rm tr}}
\newcommand{\rf}[1]{Eq.~(\ref{#1})}

\newcommand{\rfn}[1]{(\ref{#1})}

\newcommand{\rfc}[1]{Ref.~\cite{#1}}

\def\fplusrsxp{f^+_{rs}(x,p)}

\def\fminusrsxp{f^-_{rs}(x,p)}

\def\SmunuU{{\Sigma}^{\mu\nu}}

\def\S0iU{{\Sigma}^{0i}} 
 
\def\SmnU{{\Sigma}^{\mu\nu}}

\def\ubarrp{{\bar u}_r(p)}

\def\usp{u_s(p)}

\def\urp{u_r(p)}

\def\vbarsp{{\bar v}_s(p)}

\def\vrp{v_r(p)}

\def\bmu{\beta_\mu}

\def\n0{n_{(0)}}
\def\e0{\varepsilon_{(0)}}
\def\P0{P_{(0)}}

\def\omnL{\omega_{\mu\nu}}
\def\omnU{\omega^{\mu\nu}}

\def\omnUD{\tilde {\omega}^{\mu\nu}}

\def\pmu{p^\mu}

\def\Weqxk{{\cal W}_{\rm eq}(x,k)}
\def\Weqpxk{{\cal W}^{+}_{\rm eq}(x,k)}
\def\Weqmxk{{\cal W}^{-}_{\rm eq}(x,k)}
\def\Weqpmxk{{\cal W}^{\pm}_{\rm eq}(x,k)}

\def\Weqpm{{\cal W}^{\pm}_{\rm eq}}

\def\Feqpm{{\cal F}^{\pm}_{\rm eq}}

\def\P{{\cal P}}

\def\Peqpm{{\cal P}^{\pm}_{\rm eq}}

\begin{document}

\title{Kinetic approach to polarization-vorticity coupling and hydrodynamics with spin%
\thanks{Presented at the Workshop on Particle Correlations and Femtoscopy, Krakow, Poland, 22--26 May 2018}
}
\author{Avdhesh Kumar
	\address{Institute of Nuclear Physics, Polish Academy of Sciences, PL-31-342 Krakow, Poland}
}
\maketitle
\begin{abstract}
Recently introduced equilibrium Wigner functions for spin-$\onehalf$ particles are used in the semiclassical kinetic equations  to study the relation between spin polarization and vorticity. It is found, in particular, that such a framework does not necessarily imply that the thermal-vorticity and spin polarization tensors are equal. Subsequently, a procedure to formulate the hydrodynamic framework for particles with spin-$\onehalf$, based on the semiclassical expansion, is outlined.
\end{abstract}



\section{Introduction}
\label{intro}
Fireballs of strongly interacting matter formed in non-central heavy-ion collisions carry very large global angular momentum  \cite{Becattini:2007sr} which may induce spin polarization similarly to magnetomechanical effects of Einstein and de~Haas \cite{dehaas:1915} and Barnett \cite{Barnett:1935}. Therefore, the  first positive measurements of $\Lambda$--hyperon spin polarization~\cite{STAR:2017ckg,Adam:2018ivw} in heavy-ion collisions brought about a widespread interest in theoretical studies related to spin polarization and vorticity.  A natural framework that can deal simultaneously with polarization and vorticity is hydrodynamics with spin. Its relativistic variant has been recently proposed in Refs.~\cite{Florkowski:2017ruc,Florkowski:2017dyn}, see also \cite{Florkowski:2018myy}. 
 
 In this proceedings contribution we report on our recent work \cite{Florkowski:2018ahw}, where we performed a
 critical comparison of the thermodynamic and kinetic approaches which deal with spin
 polarization and vorticity. Thermodynamic approach refers to the general properties 
 of matter in global equilibrium with a rigid rotation \cite{Becattini:2013fla, Becattini:2009wh, Becattini:2012tc, Becattini:2015nva, Becattini:2013vja, Becattini:2016gvu, Becattini:2017gcx}, whereas the kinetic approach relies on the study of kinetic equations (with the vanishing collision term) for the Wigner functions of spin-$\onehalf$ particles, discussed recently in Refs.~\cite{Gao:2012ix,Chen:2012ca,Fang:2016vpj,Fang:2016uds}. We also outline herein the procedures to construct hydrodynamics with spin.

\section{Global and local equilibrium}

For spinless particles the phase space distribution function $f(x,p)$ satisfies the Boltzmann equation of the form
\beq
p^\mu \p_\mu f(x,p) = C[f(x,p)],
\label{simpkeq}
\eeq
where $C[f]$ is the collision integral. The latter vanishes for free streaming particles and in global or local equilibrium. In the global thermodynamic equilibrium the equation $p^\mu \p_\mu f_{\rm eq} = 0$ is satisfied exactly, leading to the following conditions for the hydrodynamic parameters $\xi=\mu/T$ and $\beta_\mu = u_\mu/T$ appearing in the definition of $f_{\rm eq}$: $\p_\mu \xi=0$ and $\p_\mu \beta_\nu + \p_\nu \beta_\mu = 0$. The last formula is known as the Killing equation. It has the solution of the form $\beta_\mu =  \beta^0_\mu + \varpi^0_{\mu \nu} x^\nu$, where the vector $\beta^0_\mu$ and the antisymmetric tensor $\varpi^0_{\mu \nu}$ are constants. For any form of the $\beta_\mu$ field, thermal vorticity is defined as $\varpi_{\mu \nu} = -\frac{1}{2} \left(\p_\mu \beta_\nu - \p_\nu \beta_\mu \right)$. In global equilibrium $\varpi_{\mu \nu}=\varpi^0_{\mu \nu}$, hence the thermal vorticity in global equilibrium is constant.  In local equilibrium, the equation $p^\mu \p_\mu f_{\rm eq}(x,p) = 0$ is not satisfied exactly. This is so because in this case a correction $\delta f$ should be added to the equilibrium function $f_{\rm eq}$ in order to describe dissipative effects. Nevertheless, the hydrodynamic parameters may be constrained by taking specific moments of \rf{simpkeq}. They are constructed to yield the conservation laws for charge, energy, and linear momentum. 

Treatment of particles with spin involves the Wigner functions $\Weqpmxk$ which depend additionally on the antisymmetric spin polarization tensor $\omega_{\mu\nu}$ \cite{Becattini:2018duy}. This means that we may distinguish between four rather than two different types of equilibria: 1) {\it global equilibrium} --- where $\beta_\mu$ is a Killing vector, $\varpi_{\mu \nu} = \omega_{\mu\nu}  = -\frac{1}{2} \left(\p_\mu \beta_\nu - \p_\nu \beta_\mu \right) = \hbox{const}$,  $\xi = \hbox{const}$, 2)  {\it extended global equilibrium} --- $\beta_\mu$ is a Killing vector, $\varpi_{\mu \nu} \neq \omega_{\mu\nu} = \hbox{const}$,  $\xi = \hbox{const}$, 3) {\it local equilibrium} --- $\beta_\mu$ field is not a Killing vector but $\omega_{\mu\nu}(x) = \varpi_{\mu \nu}(x)$ and $\xi = \xi(x)$, and finally 4)  {\it extended local equilibrium} --- $\beta_\mu$ field is not a Killing vector, $\omega_{\mu\nu}(x) \neq \varpi_{\mu \nu}(x)$, and $\xi = \xi(x)$. Similarly to the spinless case, the global and extended global equilibrium states correspond to the case where $\Weqxk$ satisfies exactly the collisionless kinetic equation, while the local and extended local equilibrium states correspond to the case where only certain moments of the kinetic equation for $\Weqxk$ can be set equal to zero, which results in the perfect-fluid hydrodynamics with spin. 

\section{Equilibrium Wigner functions}
 \label{sec:eqwig} 

Our considerations are based on the relations between the Wigner functions and the phase-space dependent spin density matrices  $f^\pm_{rs}(x,p)$ introduced by de Groot, van Leeuwen, and van Weert (GLW) in Ref.~\cite{deGroot:1980} (here ``$+$'' stands for particles and ``$-$'' for antiparticles).  For any of the equilibrium states defined above we use the expressions from~Ref.~\cite{Becattini:2013fla}:
 \beq
 \fplusrsxp =
 	\frac{1}{2m} \ubarrp X^+ \usp, ~~~~~~~~ \fminusrsxp=- \frac{1}{2m}\vbarsp X^- \vrp.
 \eeq	
Here $m$ denotes the particle mass, while $\urp, $ and $\vrp$ are the Dirac bispinors with spin indices $r$ and $s$ running from 1 to 2. The matrices $ X^{\pm}$ are defined by the formula $X^{\pm} =  \exp\left[\pm \xi- \bmu \pmu \right] M^\pm$, with  $M^\pm=\exp\left[\pm \f{1}{2} \omnL(x)  \SmunuU \right]$ and $\SmunuU = \f{i}{4}[\gamma^\mu,\gamma^\nu]$ being the Dirac spin operator. Following Refs.~\cite{Florkowski:2017ruc,Florkowski:2017dyn} we assume herein that the spin polarization tensor satisfies the conditions $\omnL \omnU \geq 0$ and $\omnL \omnUD = 0$, where $\omnUD=\frac{1}{2}\epsilon_{\mu\nu\alpha\beta}\omega^{\alpha\beta}$. In this case  $M^\pm = \cosh(\zeta) \pm  \f{\sinh(\zeta)}{2\zeta}  \, \omnL \SmunuU$ with $\zeta  = \f{1}{2} \sqrt{\f{1}{2} \omnL \omnU}$. The parameter $\zeta$ can be interpreted as the ratio of the spin chemical potential $\Omega$ and the temperature $T$, namely $\zeta=\Omega/T$~\cite{Florkowski:2017ruc}.

Wigner functions are 4$\times$4 matrices that can be always decomposed in terms of the $16$ independent generators of the Clifford algebra. In the equilibrium cases discussed in this section
\beq
\Weqpm &=& \f{1}{4} \left[ \Feqpm + i \gamma_5 \Peqpm + \gamma^\mu {\cal V}^\pm_{{\rm eq}, \mu} + \gamma_5 \gamma^\mu {\cal A}^\pm_{{\rm eq}, \mu}
+ \SmnU {\cal S}^\pm_{{\rm eq}, \mu \nu} \right]. \label{eq:equiwfn}
\eeq
The total Wigner function is the sum $\Weqxk = \Weqpxk + \Weqmxk$. The coefficient functions appearing in the expansion of the equilibrium Wigner function can be obtained from the 
 traces of $\Weqpmxk$ contracted first with the appropriate gamma matrices, for explicit formulas see Ref.~\cite{Florkowski:2018ahw}.   
 
\vspace{-0.25cm} 
\section{Semi-classical expansion and kinetic equations} 
\label{sec:sceke}

For an arbitrary Wigner function ${\cal W}$, its spinor decomposition has the form analogous to Eq.~(\ref{eq:equiwfn}) with the corresponding coefficient functions ${\cal F}$, ${\cal P}$, ${\cal V}_\mu$, ${\cal A}_\mu$,  and ${\cal S}_{\mu\nu}$. In the (extended) global equilibrium, the function ${\cal W}$ should satisfy exactly the equation~\cite{Vasak:1987um,Florkowski:1995ei}
\bel{eq:eqforW}
\left(\gamma_\mu K^\mu - m \right) {\cal W}(x,k) = 0,  \quad\quad     {K^\mu = k^\mu + \frac{i \hbar}{2} \,\p^\mu.}\label{eq:eqforW}
\eel
In this way one obtains the constraints on hydrodynamic variables $\mu$, $T$, $u^{\mu}$ and $\omega_{\mu\nu}$. The solution of \rf{eq:eqforW} can be 
	written in the form of a series in $\hbar$ 
\beq
{\cal X} = {\cal X}^{(0)}  + \hbar {\cal X}^{(1)}  +  \hbar^2 {\cal X}^{(2)}   + \cdots,  \quad 	{\cal X} \in \{{\cal F}, {\cal P}, {\cal V}_\mu,{\cal A}_\mu,  {\cal S}_{\nu\mu} \}.
\eeq
Including the zeroth and first orders terms of  the $\hbar$ expansion, one finds the following equations for the coefficients functions ${\cal F}_{(0)}$ and ${\cal A}^\nu_{(0)}$~\cite{Vasak:1987um,Florkowski:1995ei},
	\bel{eq:kineqF0}
	k^\mu \p_\mu {\cal F}_{(0)}(x,k) = 0, \quad k^\mu \p_\mu \, {\cal A}^\nu_{(0)} (x,k) = 0, \quad
k_\nu \,{\cal A}^\nu_{(0)} (x,k) = 0.
	\eel
The other coefficients functions ${\cal X}^{(0)}$ can be expressed in terms of ${\cal F}_{(0)}$ and ${\cal A}^\nu_{(0)}$.	It turns out that such algebraic relations are obeyed by the equilibrium coefficients ${\cal X}_{\rm eq}$, hence we may assume that ${\cal X}^{(0)}={\cal X}_{\rm eq}$. Using  ${\cal F}_{\rm eq}(x,k)$ and ${\cal A}^\nu_{\rm eq} (x,k)$ in Eqs.~\rfn{eq:kineqF0} we can check that they are exactly fulfilled if $\beta^{\mu}$ is the Killing vector, while the parameters $\xi$ and  $\omega_{\mu \nu}$ are constant,  although $\varpi_{\mu \nu}$ may be different from $\omega_{\mu\nu}$. This situation corresponds in general to the case of extended global equilibrium defined above.

\vspace{-0.25cm}
\section{Formulation of hydrodynamics with spin} 
\label{sec:formulation}

Let us now turn to the discussion of the conserved currents. We include them up to the first order in $\hbar$~\footnote{We assume that  ${\cal F}_{(1)}=0$ and ${\cal A}^\nu_{(1)}=0$, however, we include the first order corrections generated by  using ${\cal F}_{(0)}$ and ${\cal A}^\nu_{(0)}$ in the kinetic equation \rfn{eq:eqforW}. }. The charge current ${\cal N}^\alpha (x)$ can be expressed in terms of the Wigner function as the following integral~\cite{deGroot:1980}
\beq
{\cal N}^\alpha  
&=&  \tr \int d^4k \, \gamma^\alpha \, {\cal W}
=   \int d^4k \, {\cal V}^\alpha.
\label{eq:Nalphacal1}
\eeq
One finds that ${\cal N}^\alpha_{\rm eq}  = N^\alpha_{\rm eq} + \delta  N^\alpha_{\rm eq}$ and  $ \p_\alpha \, \delta N^\alpha_{\rm eq}  = 0$. Thus, the conservation law for the charge current can be expressed by the equation $\p_\alpha N^\alpha_{\rm eq}(x)  = 0$, where the charge current  $N^\alpha_{\rm eq}(x)$ agrees with that obtained in \rfc{Florkowski:2017ruc}.  

In the GLW formulation~\cite{deGroot:1980}, the energy-momentum and spin tensors  are expressed as 	%
	\bel{eq:tmunu1}
	T^{\mu\nu}_{\rm GLW}=\frac{1}{m}\tr \int d^4k \, k^{\mu }\,k^{\nu }{\cal W}=\frac{1}{m} \int d^4k \, k^{\mu }\,k^{\nu } {\cal F},
	\eel
	\vspace{-0.5cm}
		\bel{eq:Smunulambda_de_Groot1}
		S^{\lambda , \mu \nu }_{\rm GLW} =\frac{\hbar}{4} \, \int d^4k \, \tr \left[ \left( \left\{\sigma ^{\mu \nu },\gamma ^{\lambda }\right\}+\frac{2 i}{m}\left(\gamma ^{[\mu }k^{\nu ]}\gamma ^{\lambda }-\gamma ^{\lambda }\gamma ^{[\mu }k^{\nu ]}\right) \right) {\cal W} \right].
		\eel
	 Carrying out the momentum integral in \rf{eq:tmunu1} we reproduce the perfect-fluid formula for the GLW energy-momentum tensor derived earlier in \rfc{Florkowski:2017ruc}.
	It should obey the conservation law $\p_\alpha T^{\alpha\beta}_{\rm GLW}(x) = 0$. If the energy-momentum tensor is symmetric, the conservation of orbital and spin parts of the total angular momentum holds separately, so we also have $\p_\lambda S^{\lambda , \mu \nu }_{\rm GLW}(x) = 0 $.
	
The canonical forms of the energy-momentum and spin  tensors, $T^{\mu\nu}_{\rm can}(x)$ and $S^{\lambda , \mu \nu }_{\rm can}(x)$, can be obtained from the Dirac Lagrangian by applying the Noether theorem and are given by the formulas
\bel{eq:tmunu1can1}
T^{\mu\nu}_{\rm can}= \int d^4k \,k^{\nu } {\cal V}^\mu,
\eel
\vspace{-0.75cm}
\beq
S^{\lambda , \mu \nu }_{\rm can}&=& \frac{\hbar}{4} \, \int d^4k \,\text{tr}\left[ \left\{\sigma ^{\mu \nu },\gamma ^{\lambda }\right\} {\cal W}  \right] = \frac{\hbar}{2} \epsilon^{\kappa \lambda \mu \nu} \int d^4k \, {\cal A}_{ \kappa}.
\label{eq:Smunulambda_canonical1}
\eeq
One can check that $T^{\mu\nu}_{\rm can} = T^{\mu\nu}_{\rm GLW} + \delta T^{\mu\nu}_{\rm can}$,  where $\delta T^{\mu\nu}_{\rm can}  = -\partial_\lambda S^{\nu , \lambda \mu }_{\rm GLW}$.  The canonical energy-momentum tensor should be conserved as well, hence we demand that  $\p_\alpha T^{\alpha\beta}_{\rm can}(x) = 0$. Since $S^{\nu , \lambda \mu }_{\rm GLW}(x)$ is antisymmetric in the indices $\lambda$ and $\mu$, we find that $\partial_\mu \, \delta T^{\mu\nu}_{\rm can}(x) = 0$. Thus, the conservation law for energy and momentum in the canonical case is reduced to the same formula as that obtained in the GLW case.

The canonical equilibrium spin tensor can be obtained by considering the axial-vector component in \rf{eq:Smunulambda_canonical1} in the zeroth order. Assuming $ {\cal A}^{(0)}_{ \kappa}= {\cal A}_{\rm eq,  \kappa}$ and carrying out the integration over $k$ one gets $S^{\lambda , \mu \nu }_{\rm can}  =S^{\lambda , \mu \nu }_{\rm GLW} + S^{\mu , \nu \lambda }_{\rm GLW}+ S^{\nu , \lambda \mu }_{\rm GLW}$. One can show that  $\p_\lambda S^{\lambda , \mu \nu }_{\rm can}(x)  = T^{\nu\mu}_{\rm can} - T^{\mu\nu}_{\rm can}$. This is an interesting result as one can see  that the energy-momentum tensor is not symmetric in the canonical case. It is important to note that the two approaches (GLW and Canonical) are connected via a pseudo gauge transformation~\cite{Florkowski:2018ahw}. 

Conservation laws for charge, energy, and momentum can be obtained by taking the zeroth and first moments of the kinetic equation $k^\mu \p_\mu {\cal F}_{(0)} = 0$. Since we have an additional degree of freedom connected with spin polarization, the equations for charge and energy-momentum are not closed. In oder to close them one needs to determine the dynamics of spin which can be obtained by multiplying the kinetic equation for the axial coefficient of the Wigner function  \rfn{eq:kineqF0} by a factor $\epsilon^{\mu\beta\gamma\delta}k_{\beta}$ and then by integrating over $k$. In this way, we obtain the conservation of the GLW version of the spin tensor.  

\vspace{-0.5cm}
\section{Summary and conclusions} \label{sec:summary}
We have introduced the Wigner functions using the equilibrium distribution functions of particles with spin-$\onehalf$ put forward in Ref.~\cite{Becattini:2013fla}. Using kinetic equations for the Wigner function we have found that  the kinetic approach  does not necessarily imply a direct relation between the thermal vorticity and spin polarization, except for the fact that the two should be constant in global equilibrium.  We have furthermore outlined the procedures to construct the hydrodynamic equations with spin. 

	This work was supported in part by the Polish National Science Center Grant No. 2016/23/B/ST2/00717.

%
\end{document}